\documentclass[aps,pra,twocolumn,showpacs,superscriptaddress]{revtex4-1}

\usepackage{amsmath}
\usepackage{amssymb}
\usepackage{amsfonts}
\usepackage{txfonts}
\usepackage{graphicx}
\usepackage{hyperref}
\newcommand {\eref}[1]{(\ref{#1})}

\newcommand{\rr}{\mathbf{r}}
\newcommand{\vn}{{\rm v}}

\begin{document}
\title{Classical and quantum regimes of two-dimensional turbulence\\ in trapped Bose-Einstein condensates}
\author{M. T. Reeves}
\affiliation{Jack Dodd Center for Quantum Technology, Department of Physics, University of Otago, Dunedin, New Zealand.}
\author{B. P. Anderson}
\affiliation{College of Optical Sciences, University of Arizona, Tucson, Arizona 85721, USA.}
\author{A. S. Bradley}
\affiliation{Jack Dodd Center for Quantum Technology, Department of Physics, University of Otago, Dunedin, New Zealand.}

\date{\today}
\begin{abstract}
We investigate two-dimensional  turbulence in  finite-temperature trapped Bose-Einstein condensates within damped Gross-Pitaevskii theory. Turbulence is produced via circular motion of a Gaussian potential barrier  stirring the condensate. We systematically explore a range of stirring parameters and identify three regimes, characterized by the injection of distinct quantum vortex structures into the condensate:  (A) periodic vortex dipole injection, (B) irregular injection of a mixture of vortex dipoles and co-rotating vortex clusters, and (C) continuous  injection of oblique solitons that decay into vortex dipoles. Spectral analysis of the kinetic energy associated with vortices reveals that regime (B)  can intermittently exhibit a Kolmogorov $k^{-5/3}$ power law over almost a decade of length or wavenumber ($k$) scales. The kinetic energy spectrum of regime (C) exhibits a clear $k^{-3/2}$ power law associated with an inertial range for weak-wave turbulence, and a $k^{-7/2}$ power law for high wavenumbers. We thus identify distinct regimes of forcing for generating either two-dimensional quantum turbulence or classical weak-wave turbulence that may be realizable experimentally.
\end{abstract}
\maketitle

\section{Introduction}
Quantum turbulence (QT)~\cite{Vinen07a} involves chaotic vortex dynamics in a superfluid~\cite{Bar2001.book.Superfluid,Berloff02a,Tsubota09a,Kozik09a,Nowak2011a} and is often associated with a random vortex tangle in three dimensions (3D)~\cite{Bar2001.book.Superfluid}. QT experiments have been conducted for over half a century, and initially experiments were performed using superfluid helium~\cite{Bar2001.book.Superfluid}.  More recently, attention has turned to superfluid Bose-Einstein condensates (BECs); these lend themselves well to the problem, as many condensate parameters can be finely controlled~\cite{Tsubota09a}. Moreover, modern optical techniques routinely allow visualization of vortex cores in ballistically expanded BECs, whereas vortex visualization is challenging in superfluid helium, although possible~\cite{Bewley08b}. The presence of the Kolmogorov spectrum~\cite{Kolmogorov1941} has been established in superfluid helium 3DQT, both in experiments \cite{Maurer1998a} and  quantized vortex filament model simulations \cite{Araki2002}. Numerical studies of 3DQT in BECs using the Gross-Pitaevskii equation (GPE) have also established the presence of a Kolmogorov spectrum~\cite{Nore97a,Kobayashi05a,Kobayashi2007a,Kobayashi2008}. Such evidence has aroused strong interest in the similarities between classical turbulence (CT) and QT, and it is speculated that studies of QT may help progress the classical theory~\cite{Barenghi08a}. 
\par
In forced 3D turbulence, a classical fluid exhibits a direct cascade of energy from the forcing scale down to the damping scale.  This sets the size of the inertial range~\cite{Sreenivasan1999a} over which the kinetic energy spectrum approximates the Kolmogorov $k^{-5/3}$ law over wavenumbers $k$.   
Incompressible two-dimensional (2D) classical fluids exhibit very different flow characteristics due to the existence of an additional inviscid invariant, namely the total squared vorticity, or {\em enstrophy}~\cite{Kra1967.PF10.1417,Lei1968.PF11.671,Bat1969.PF12.II233,Kra1980.RPP5.547}. Consequently small-scale forcing induces vorticity to aggregate into coherent rotating structures~\cite{Kellay2002a}, associated with an \emph{inverse} cascade of energy to progressively larger length scales~\cite{Mon1974.PF6.1139}.  In a distinct range of scale-space, a direct enstrophy cascade occurs, in which enstrophy is conservatively transported from the forcing scale to progressively smaller scales~\cite{Les2008.Turbulence}.  Thus in 2D classical turbulence, the kinetic energy spectrum can exhibit the $k^{-5/3}$ power law in the inertial range, and a $k^{-3}$ power law in the scale range of the enstrophy cascade~\cite{Kra1967.PF10.1417}. The inverse-cascade phenomenon has been widely studied in classical fluids~\cite{Les2008.Turbulence,Boffetta12a}, and the dual-cascade spectrum has been observed in experiments with soap films~\cite{Rutgers1998}.
\par~
A great deal is known about 3DQT in BECs~\cite{Berloff02a,Kobayashi05a,Proment09a,Kozik09a}, and experiments have observed \cite{Raman2001,Sch2004.PRL93.210403} and explicitly studied \cite{Henn09a,Seman2011a,Car2012.JLTP2012.49} characteristics of 3DQT in BECs.  Much less is known about 2DQT; more work is required to understand the fundamental characteristics of 2DQT, and to compare the classical and quantum dynamics. Experimental progress on 2DQT in BECs has concentrated on methods to generate disordered vortex distributions in highly oblate condensates and observe the decay of these turbulent states \cite{Neely2012a,TWNThesis,ECSThesis}. Despite growing theoretical interest in 2DQT~\cite{Chu2001a,Chu2001b,Parker05a,Nazarenko06a,Wang2007a,Numasato2009,Horng09a,Numasato10a,Sasaki2010,White10a,Numasato10b,Schole2012a,Bradley2012a,White2012a}, work in BEC has largely focused on decaying turbulence~\cite{Horng09a,Numasato10a,Schole2012a}, where vortex-antivortex recombination may generate a direct energy cascade~\cite{Numasato10b}. However, if forcing can be chosen to generate sufficient clustering of vortices of the same sign of circulation~\cite{Sasaki2010,Bradley2012a,White2012a,Neely2012a}, recombination can be suppressed, and an inverse-energy cascade may be possible~\cite{Bradley2012a,Neely2012a}. Indeed, recent large-scale numerical modeling of an experiment involving a highly oblate 3D system with effective 2D vortex confinement exhibited characteristics consistent with an inverse energy cascade~\cite{Neely2012a}. 
\par
In regimes dominated by acoustic radiation, a different type of turbulence known as {\em weak-wave} turbulence (WWT)~\cite{Zakharov1992,Dyachenko1992} can occur. WWT is a classical wave phenomena that has been studied in Gross-Pitaevskii theory~\cite{Berloff02a,Nazarenko06a,Proment09a,Kozik09a,Nowak2011a}. In the low temperature regime the existence of a large BEC causes the nonlinear interactions to be dominated by three-wave processes which lead to characteristic power laws in the kinetic energy spectrum~\cite{Nazarenko06a}. In particular, three-wave 2D WWT in BECs is predicted to generate a $k^{-3/2}$ spectrum at long wavelengths, associated with a direct cascade of wave energy~\cite{Nazarenko06a}. In general, the power laws observed in a particular scale range may depend on the effectiveness of damping at that scale.
\par
Here we consider turbulent flows generated by forced stirring of an oblately confined BEC with a laser-generated Gaussian potential. Stirring an oblately confined BEC in this way can excite the superfluid into highly disordered states suggestive of turbulence. Focusing on the distinction between 2DQT and 2DWWT, we investigate the relationship between the stirring characteristics and the kind of excited flow states generated. We systematically study a range of experimentally accessible stirring parameters, and classify the resulting superfluid dynamics through analysis of kinetic energy spectra and vortex clustering dynamics. 
\par
This paper is structured as follows. In Section \ref{sec:model} we discuss our model. In Section \ref{sec:regimes} we discuss our choice of simulation parameters, and identify distinct vortex injection regimes within the range of stirring parameters considered. In Section \ref{sec:spectra} we qualitatively analyze the kinetic energy spectra and  kinetic energy composition of a characteristic example from each vortex injection regime. In Section \ref{sec:fit} we characterize power law behavior and intermittency of spectra using linear least-squares fit analysis. In Section \ref{sec:D&C} we discuss our results and conclude.

\section{Damped Gross-Pitaeveskii equation}
\label{sec:model}
The Hamiltonian for a three-dimensional Bose gas described by field operator $\hat{\psi}(\mathbf{r},t)$ is
\begin{equation}\label{fullH}
H=\int d^3\mathbf{r}\;\hat{\psi}^\dag(\mathbf{r},t)H_{sp}\hat{\psi}(\mathbf{r},t)+\frac{g}{2}\hat{\psi}^\dag(\mathbf{r},t)\hat{\psi}^\dag(\mathbf{r},t)\hat{\psi}(\mathbf{r},t)\hat{\psi}(\mathbf{r},t),
\end{equation}
with single-particle Hamiltonian $H_{sp}=-\hbar^2\nabla^2/2m+V(\mathbf{r},t)$, trapping potential $V(\mathbf{r},t)$, atomic mass $m$, interaction parameter $g=4\pi\hbar^2a/m$, and $s$-wave scattering length $a$. We consider confinement by a cylindrically symmetric harmonic trap, $V_{ho}(\rr)$, augmented by a Gaussian stirring potential, $V_{s}(\rr,t)$. The resulting trapping potential is 
\begin{equation}
V(\rr,t)=V_{ho}(\rr)+V_s(\rr,t)
\end{equation}
where $V_{ho}(\rr)=m\omega_r^2(x^2+y^2)/2+m\omega_z^2z^2/2$ for trapping frequencies $\omega_r$, $\omega_z$. We consider the case of strong confinement $\hbar \omega_z \gg \mu, k_BT, \hbar \omega_r$, where $\mu$ is the chemical potential and $T$ is the system temperature. Under these circumstances, the condensate adopts a highly oblate `pancake' shape, and is effectively two-dimensional. The effective interaction parameter in 2D is $g_{\rm{2D}} = g/\sqrt{2\pi} \ell_z$ where $\ell_z = \sqrt{\hbar/m\omega_z}$ is the harmonic oscillator length in the $z$ direction. We note that our study of this highly oblate 2D system has wider applicability to less oblate systems which also exhibit effective 2D vortex dynamics~\cite{Neely10a,Rooney11a}.

Treatment of \eref{fullH} within a detailed reservoir interaction theory leads to the Stochastic Projected Gross-Pitaevskii equation (SPGPE)~\cite{Gardiner03a}, which describes the evolution of atoms with energy less than a chosen cutoff energy and their interaction with thermalized atoms above the cutoff. The description we use can be obtained from the simple growth SPGPE~\cite{Bradley08a} by neglecting the thermal noise and retaining the damping term.
This leads to the damped Gross-Pitaevskii equation (dGPE)
 \begin{equation}
i\hbar \frac{\partial\psi(\rr,t)}{\partial t} = L\,\psi(\rr,t) + i\gamma\left[\mu - L\right]\psi(\rr,t),
\label{eq:DGPE}
\end{equation}
describing the purely dissipative evolution of the condensate wavefunction $\psi(\mathbf{r},t)$ due to a stationary thermal reservoir.
The operator $L$ is given by
\begin{equation}
L\,\psi(\rr,t) \equiv \left[- \frac{\hbar^2\nabla_{\perp}^2}{2m} + V(\rr,t) + g_{\rm{2D}}|\psi(\rr,t)|^2 \right]\psi(\rr,t).
\end{equation}
The damping $\gamma$ can be derived ab-inito for quasi-equilibrium states, and it is typically very small (of order $10^{-4}$)~\cite{Blakie08a}.
In this work we take the approach of modeling an experimentally realizable system in all respects apart from the damping, where we neglect the details of a full finite-temperature theory. Instead we treat the dimensionless damping parameter $\gamma$ phenomenologically, choosing $\gamma$ to be much smaller than all other dimensionless rates governing the dynamics. 

We stir the superfluid by introducing a time-dependent repulsive Gaussian potential, which represents a blue-detuned laser beam propagating along $z$, of the form
\begin{equation}
\label{TheLaser}
V_s(x,y,t) = V_0 \exp \left[- \frac{(x - x_0(t))^2 + (y-y_0(t))^2}{\sigma^2}\right],
\end{equation}
where $(x_0(t),y_0(t))$ specifies the location of the stirring beam center.
Ground states are obtained by propagating (\ref{eq:DGPE}) for $\gamma \equiv 1$ using the Thomas-Fermi wavefunction as an initial condition, with the stationary Gaussian obstacle  (\ref{TheLaser}) at its initial position $(x_0(0), y_0(0)) = (s,0)$. Our choice of stirring procedure allows for several parameters to be varied. The energy required to form a vortex dipole is a minimum at approximately $ s = 0.4R_{\rm{TF}}$ in the Thomas-Fermi regime~\cite{Zhou2004}, where $R_{TF}$ is the Thomas-Fermi radius in the radial dimension, and there is experimental evidence in agreement with this prediction \cite{Raman2001}. We therefore consider only circular stirring symmetric about the trap center, such that $x_0 = s\cos({\textrm v}t/s)$ and $y_0 = s \sin({\textrm v}t/s)$, where $\textrm{v}$ is the speed of the stirrer.  

We work in units of energy, length and time given by $\mu$, $\xi$ and $\xi/c$ respectively, where $\xi$ is the healing length ($\hbar^2/m\xi^2\equiv\mu$) and $c=\sqrt{\mu/m}$ is the speed of sound. The integration routine we implement is a pseudo-spectral adaptive Runge-Kutta method of orders 4 and 5 \cite{NR}. We choose a damping parameter of $\gamma= 0.03$, which is smaller than any other simulation parameter by at least an order of magnitude. The $1/e$ Gaussian half-width of the stirrer is chosen as $\sigma = 4\xi$ in all simulations.  

\section{Regimes of turbulence}\label{sec:regimes}
\subsection{System and parameters}
We initially performed a systematic sequence of 130 simulations over a range of obstacle strengths ($V_0$) and speeds $(\vn)$ in order to determine temporal characteristics of the vortex emission. The primary motivation for carrying out this procedure was to identify stirring parameters which are most efficient for the production of like-charge vortex clustering, i.e. turbulent behavior in the context of quantum vortex turbulence \cite{Bradley2012a}. 

 Due to the number of simulations required we choose a relatively small system, as this allows one to use a numerical grid with fewer points while still maintaining adequate spatial resolution. We choose harmonic trapping frequencies of $(\omega_r, \omega_z) = 2\pi \times (39,5000)$ Hz, so that the system is well within the 2D regime. Working with $^{87}$Rb gives a 2D interaction parameter of ${g}_{\rm{2D}} = 0.19\mu\xi^2$. Choosing a peak density of  $n_0 = 5\times 10^9{\rm cm}^{-2}$ results in a condensate with a Thomas-Fermi radius of  $R_{\rm{TF}} = 40\xi$, containing $N \approx 1.3 \times 10^4$  atoms. Values for the chemical potential, healing length and speed of sound are $\mu/k_{\rm{B}} = 53.2$nK, $\xi = 0.324 \mu$m and $c = 2.26$mm/s respectively.  For $\sigma = 4\xi$ the corresponding $1/e^2$ radius of a stirring beam would be $4\sqrt{2}\xi \sim 1.8 \mu$m, an experimentally realizable beam size. The system is simulated using a spatial domain of $L^2 = (130\xi)^2$, and a grid of  $M^2 = 512^2$ points. We have verified that the numerics are convergent for the chosen grid by testing the most violent cases on a finer grid and verifying the phenomenology.
 %============================================================== 
\begin{figure}[!t]
\begin{center}
\includegraphics[width=\columnwidth]{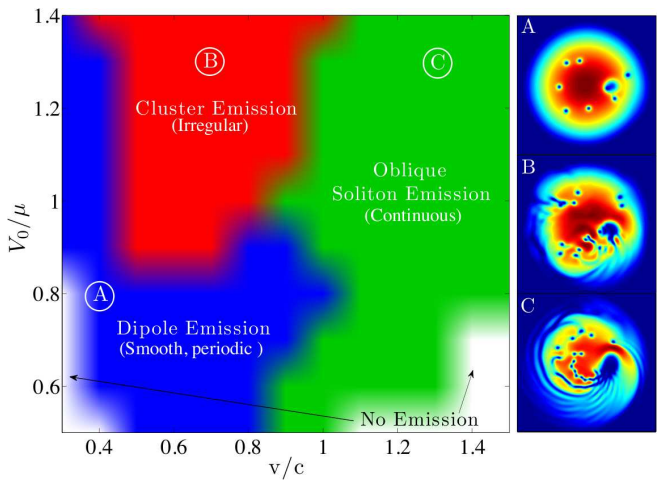}
\caption{Regimes of vortex emission (left), determined by simulating the dGPE for parameters $V_0/\mu = \{0.5,0.6, ... ,1.4\}$ and $ \textrm{v}/c = \{0.3,0.4,...,1.5\}$. The points indicated by circles are (A) $V_0/\mu = 0.8$, $\textrm{v}/c = 0.4$, (B) $V_0/\mu = 1.3$, $\textrm{v}/c = 0.7$ and (C) $V_0/\mu = 1.3$, $\textrm{v}/c = 1.3$. Blurred regions indicate the approximate extent of the transition regions. False color images of the condensate density profiles (right) display regions of low density (blue) and high density (red). The three examples A, B and C correspond to the stirring parameters indicated on the regime map.  
In each case the stirrer has completed a cycle of $2\pi$ radians. The field of view in the density profiles is  $(90\xi)^2$.}
\label{fig1}
\end{center}
\end{figure}
%============================================================== 

\subsection{Stirring Regimes}\label{sec:regimesA}
Three distinct regimes of vortex emission were found for the parameters considered. Our findings are presented in Figure \ref{fig1} as a false color map for a range of potential strengths ($V_0$) and speeds ($\vn$), accompanied by a specific example of the condensate density profile for each regime. Changes in the stirring phenomenology occur when the parameters $\vn/c$ and  $V_0/\mu$ are of order unity, corresponding respectively to the transition from subsonic to supersonic stirring speeds and from a {\em penetrable} ($V_0<\mu$) to an {\em impenetrable} ($V_0>\mu$) obstacle beam. Note that the boundaries between regions in Figure \ref{fig1} do not correspond to abrupt transitions, as there is a gradual cross-over between different regimes.

\textbf{Dipole Regime:} Above a minimum velocity, $\vn \simeq 0.3c$, single dipoles shed from the obstacle in a regular, periodic fashion. The emission is associated with a density minimum that drops to zero at the time of dipole shedding.
The dipoles occasionally interact with each other, sometimes exchanging constituent vortices, and eventually disappear in vortex-antivortex annihilations. The overall dynamics are very temporally regular, suggestive of a laminar regime. We do not observe any clusters of vortices with the same circulation being emitted from the obstacle in this regime. [Figure~\ref{fig1}, panel A].
\par
\textbf{Cluster Regime:} If we maintain a stirring velocity in the range $0.3c\lesssim \vn\lesssim c$, but increase the obstacle strength such that it becomes impenetrable, we observe that the temporal characteristics of vortex emission become increasingly irregular with increasing obstacle strength. Furthermore, we observe that some vortices which shed from the obstacle cluster with like-charged vortices. As we further increase the strength of the obstacle, the range of velocities for which we observe this behavior extends, and clustering of like-charged vortices becomes more prominent [Figure~\ref{fig1}, panel B].
\par
\textbf{Oblique Soliton Regime:} For both penetrable and impenetrable obstacles, increasing the stir velocity into the supersonic regime $ \vn>c$ causes the obstacle to shed oblique dark solitons that are unstable to decay via the snake instability into chains of vortex dipoles. A large compressional wave can also be seen in front of the obstacle. The vortex dipoles that form due to the snake instability have a small dipole length and rapidly annihilate, generating bursts of acoustic energy. Numerically we find that almost all vortices (at least $\sim 89\%$) are bound into vortex-antivortex pairs throughout the simulation, and thus we do not observe significant clustering of like-charged vortices in this regime~[Figure~\ref{fig1}, panel C]. 
\par
\textbf{Zero Emission Regime:} The white regions of the parameter map in Figure \ref{fig1} indicate the parameters for which we observe no vortex emission. It is already known that for an obstacle moving through the condensate there is a critical velocity below which vortex emission does not occur~\cite{Frisch92a,Neely10a}. However, we also find that once the speed of a penetrable obstacle is increased past an upper critical value, vortices no longer nucleate inside the BEC. Instead, surface waves are generated and vortices eventually nucleate at the condensate boundary. The density minimum dragged behind the obstacle falls further behind as its speed increases, eventually reaching a trailing distance of order the system size. Note that for sufficiently rapid stirring the BEC cannot respond to the obstacle potential and will only see a time-averaged potential. For smaller $V_0$ this regime will be reached at lower stirring speeds, a behavior that is consistent with the boundary seen in Fig \ref{fig1}, where it approaches $c$ in the regime $V_0\ll \mu$. We have also verified that this behavior occurs in the absence of dissipation.
%============================================================== 
\begin{figure}[!ht]
\includegraphics[width=\columnwidth]{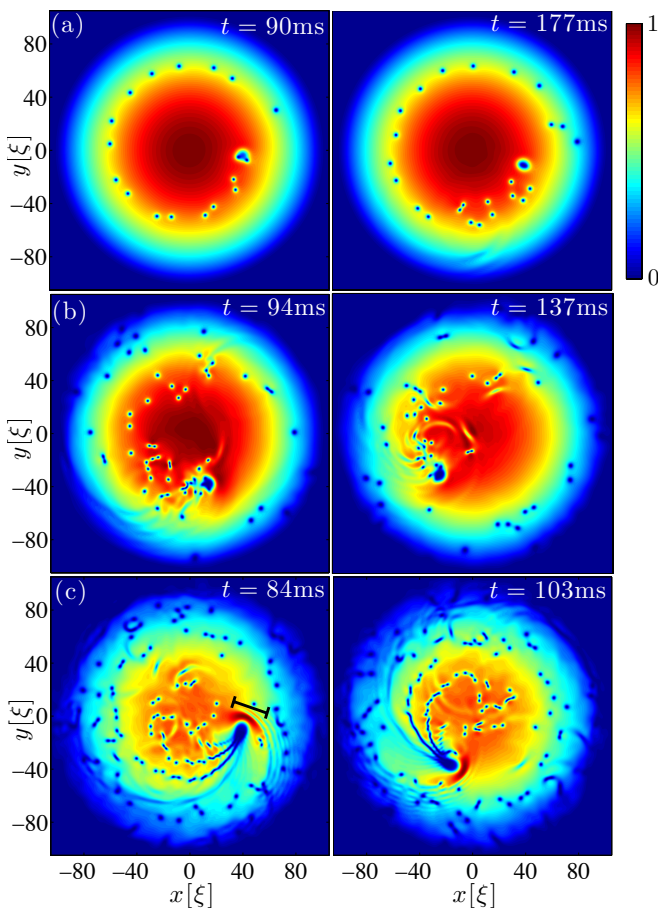} 
\caption{Density profiles produced within the (a) Dipole regime, (b) Cluster regime and (c) Oblique soliton regime, in the larger system that we have used for spectral analysis. The spectra corresponding to each density profile are shown in Figures \ref{fig3} and \ref{fig4}. The field of view is $(210\xi)^2$.  The color bar represents atomic density as a fraction of the peak density. The black bar in the bottom left panel is $\sim 30\xi$ long and indicates a  forcing source of compressible energy, as discussed in Sec.~\ref{sec:forcing}.}
\label{fig2}
\end{figure}
%============================================================== 

\section{Kinetic Energy Spectra and Energy Composition}
\label{sec:spectra}
We now consider a more detailed analysis of the points A, B, and C shown in Figure \ref{fig1}, corresponding to $\textrm{v}/c=0.4, 0.6, 1.3$ and $V_0/\mu=0.8, 1.3, 1.3$ respectively.
For the purposes of spectral analysis it is desirable to consider a system with a much greater spatial extent than that used for the previous investigation, as one typically wishes to characterize spectra over at least a decade of wavenumbers. We therefore extend the spatial domain to $L^2 = (270\xi)^2$ and reduce the radial trapping frequency to $\omega_r = 2\pi \times 16$ Hz and keep $\mu$ constant, which results in a condensate containing $N \approx 7.9 \times 10^4$ atoms with a Thomas-Fermi radius of $R_{\rm{TF}} = 100\xi$. We also increase the grid resolution to $M^2 = 2048^2$. The axial trapping frequency $\omega_z$ and the peak density $n_0$ are left unchanged, thus preserving the interaction parameter $g_{\rm{2D}}$, and healing length $\xi$. This ensures that the characterization of the previous section remains valid for this system.
For each set of parameters we evolve the system according to the dGPE \eref{eq:DGPE}, for several complete cycles of the stirring. In general the kinetic energy spectra, which we present below, fluctuate with time. To illustrate the degree of variability we choose two representative times during the motion, for which the atomic densities are shown in Figure~\ref{fig2}. In the Supplemental Material we provide movies of the dynamics in the three regimes, showing particle density, compressible and incompressible energy spectra, and the vortex distribution for the entire time evolution.

\subsection{Incompressible Kinetic Energy Spectra }\label{sec:enspecA}
\begin{figure}[!t]
\includegraphics[width=\columnwidth]{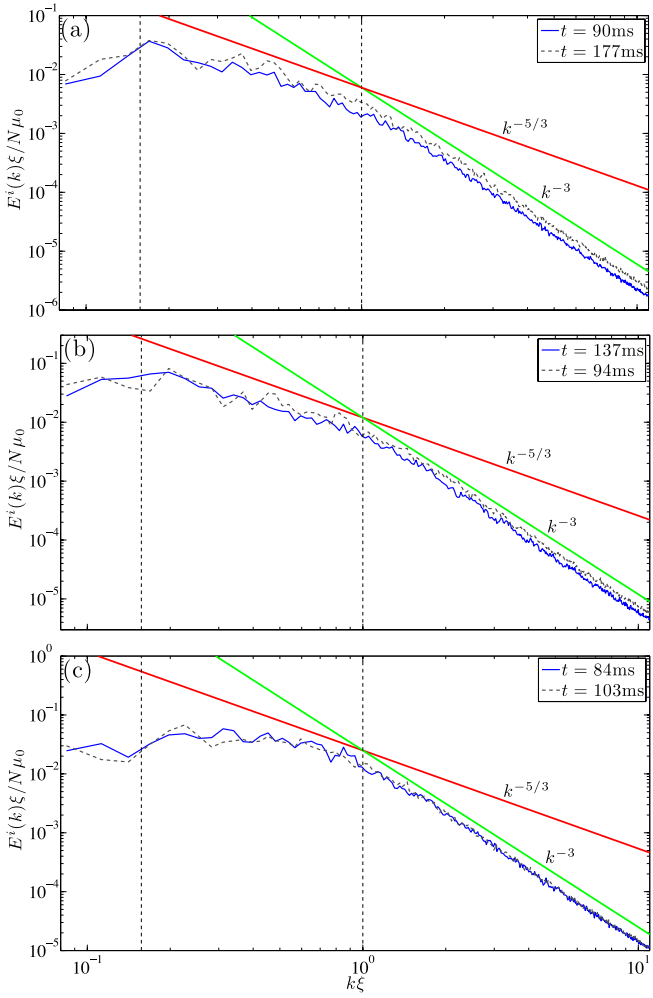} 
\caption{Scaled incompressible kinetic energy spectra produced in the (a) Dipole regime, (b) Cluster regime, and (c) Oblique soliton regime. The spectra are presented for the times shown in Figure~\ref{fig2}. Lines proportional to $k^{-5/3}$ (red) and $k^{-3}$ (green) are also shown. The vertical dashed lines are located at $k\xi = 2\pi/40$ and $k\xi = 1$. Movies displaying the temporal evolution of each spectrum are provided in the Supplemental Material.}
\label{fig3}
\end{figure}

We decompose the system energy and calculate the incompressible and compressible kinetic energy spectra according to the method outlined in Refs. \cite{Nore97a,Numasato10b}. The incompressible kinetic energy spectrum is associated with quantum vortices, and the compressible part is associated with acoustic waves. The incompressible kinetic energy spectra corresponding to the densities of Figure~\ref{fig2} are shown in Figure~\ref{fig3}.

%============================================================== 
%============================================================== 
All three regimes clearly display a $k^{-3}$ power law in the ultraviolet (UV) region ($k\xi \gg 1$) of the spectrum. This power law is robust throughout the simulation in all regimes. This is attributed solely to the internal structure of the vortex core and has no clear correspondence with a direct enstrophy cascade~\cite{Bradley2012a}. We will hence focus on the infrared (IR) region $(k\xi < 1)$, which is related to the spatial configuration of vortices~\cite{Bradley2012a}.

We observe that the spectrum of the dipole regime does resemble the Kolmogorov $k^{-5/3}$ law in the infrared region, over the scale range $2\pi/40<k\xi<1$, although some minor oscillations are observed [Figure \ref{fig3}(a), $t = 90$ms]. This is surprising due to the regular periodic nature of the vortex emission dynamics  [see Supplemental Movie 1] and suggests caution is necessary when interpreting spectra as signatures of turbulence. The spectrum does at times lose much of this resemblance, as demonstrated by the spectrum at $t = 177$ms, but in general bears some comparison to the -5/3 law while not being particularly linear in log-space.

Turning now to the cluster regime [Figure \ref{fig3}(b)],  we find that the spectrum shown in the Figure at $t=137$ms displays a close resemblance to the Kolmogorov $k^{-5/3}$ power law within the range $ 2\pi/40 < k\xi <1$. This spectral resemblance is analyzed further in Sec.~\ref{sec:fit}.  We note that $k\xi = 2\pi/40$ is the wavenumber that corresponds to the radial obstacle location.  We do however find that the power law is highly temporally intermittent, sustaining briefly but also undergoing significant distortions several times throughout the course of the simulation [see Supplemental Movie 2]. A qualitative indication of the extent to which the spectrum deviates from the power law is displayed by the spectrum at $t = 94$ms of Figure~\ref{fig3}(b).  

Figure \ref{fig3}(c) displays example spectra for the oblique soliton emission regime. This regime does not exhibit any significant clustering of like-sign vortices, and the spectrum clearly does not conform to a power law in the infrared region. The examples presented are typical of what is observed throughout the simulation. The time evolution of the power-law exponents of the dipole and cluster regime is further analyzed in Sec.~\ref{sec:fit}.

\subsection{Compressible Kinetic Energy Spectra}
%============================================================== 
\begin{figure}[!t]
\includegraphics[width = \columnwidth]{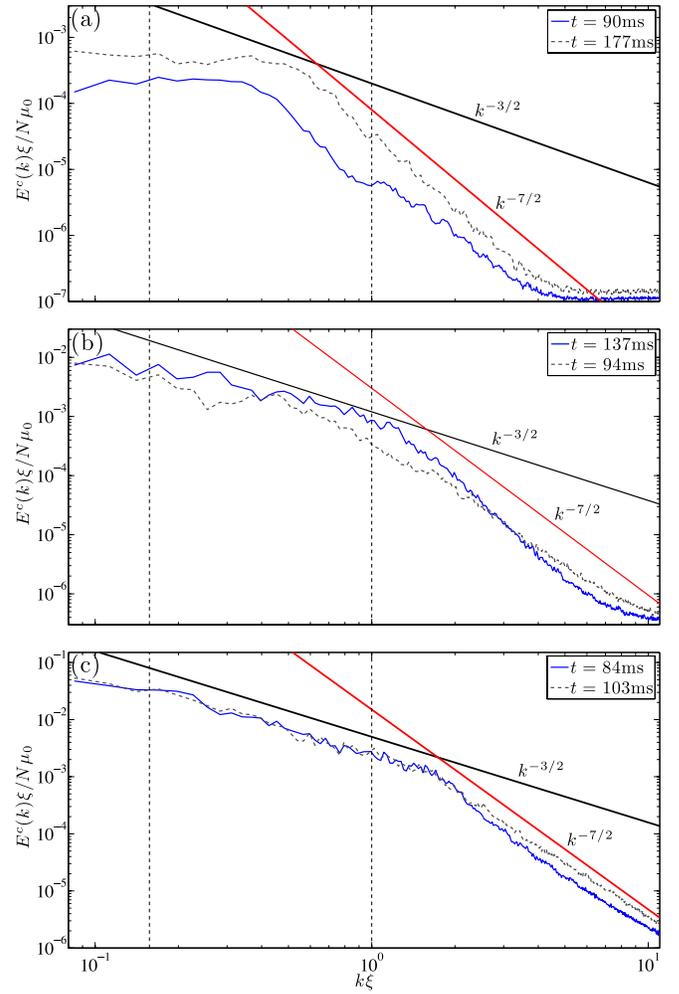} 
\caption{Scaled compressible kinetic energy spectra for (a) the Dipole regime, (b) the Clustering regime and (c) the Oblique soliton regime. The spectra are presented for the times shown in Figure~\ref{fig2}. Lines proportional to $k^{-3/2}$ (black) and $k^{-7/2}$ (red) are provided for comparison. The vertical dashed lines are located at $k\xi = 2\pi/40$ and $k\xi = 1$. Movies displaying the temporal evolution of each spectrum are provided in the Supplemental Material.}
\label{fig4}
\end{figure}
%============================================================== 

We now  describe the compressible kinetic energy spectra for our characteristic cases, shown in Figure \ref{fig4}. The dipole regime [Figure \ref{fig4}(a)] can transiently resemble power-law behavior in the UV region of the spectrum (e.g. at $t=177$ms in Figure~\ref{fig4}). This behavior is clearest during dipole annihilation events [see Supplemental Movie 1], but is seen only briefly as individual sound pulses are emitted. The spectrum quickly returns to non power-law behavior, such as that seen in Figure~\ref{fig4} at $t=90$ms. The cluster regime [Figure \ref{fig4}(b), $t =94$ms] can display power-law exponents of approximately $-3/2$ in the IR region and $-7/2$ in the UV region, but is also susceptible to significant deviations (e.g. at $t = 137$ms).  

The oblique soliton regime displays very different behavior to the other two cases. The spectrum produced in this regime [Figure \ref{fig4}(c)] displays a clear bilinear form, again with power-law exponents of $-3/2$ and $-7/2$. The $k^{-3/2}$ power law evident across a decade of wavenumbers in the IR region is consistent with an inertial range for three-wave WWT in 2D~\cite{Zakharov1992,Dyachenko1992}. The shape of the spectrum in the IR region is found to be extremely robust throughout the simulation,  as shown here by the two example spectra. The UV region exhibits larger fluctuations than the IR region, but  these fluctuations appear to be closely centered about a $-7/2$ power law. The time evolution of the power-law exponents of the clustering and oblique soliton regimes is further analyzed in Sec.~\ref{sec:fit}.

%===============================================================
 \begin{figure}[!t]
\includegraphics[width = \columnwidth]{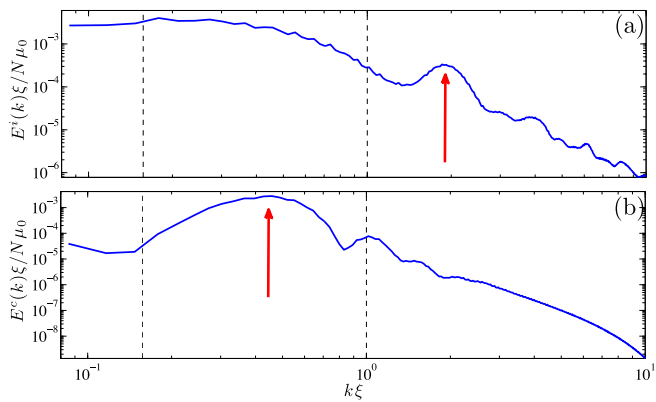} 
\caption{(a) Forcing peak at $k\xi \simeq 2$ in the incompressible energy spectrum of the soliton regime at $t = 2.5$ ms (b) Forcing peak at $k\xi \simeq 0.5$ in the compressible energy spectrum of the soliton regime at $t = 0.7$ ms.  The vertical dashed lines are located at $k\xi = 2\pi/40$ and $k\xi = 1$.}
\label{fig5c}
\end{figure}
%===============================================================

\subsection{Forcing Scales}\label{sec:forcing}
Here we identify possible sources of forcing from observing spectra at early evolution times. In the dipole and cluster regimes the incompressible spectra show no indication of a localized forcing peak. However, in the soliton regime we observe a peak localized at $k\xi \simeq 2$ [Figure \ref{fig5c}(a)]. This is consistent with rapid nucleation of many regularly spaced vortices in the early wake of the obstacle [see Supplemental movie 3]. In the other two regimes, the vortex emission is slower, and may cause the appearance of a forcing peak to be washed out.

The compressible spectrum also shows a clear source of forcing at $k\xi \simeq 0.5$, as displayed in Figure \ref{fig5c}(b). This feature is observed in all regimes, although it is largest in the soliton regime. We attribute this peak to the compression wave that forms in front of the obstacle. As time progresses, we observe that the peak drifts towards lower wavenumbers until the compression wave has developed to its full spatial extent of approximately $30\xi$, as indicated by the scale bar in Figure \ref{fig2} (c). We have also examined the compressible energy density in position space and verified that there is a high concentration of compressible energy in this region.

The distinct kink in Figure~\ref{fig4}(c) at $k\xi \simeq 2$ suggests there may be a second forcing peak at this scale. There is some indication of a feature at this scale in the compressible spectrum between $2-6$ ms [see Supplemental Movie 3], but the peak is not as prominent as those shown in Figure \ref{fig5c}. Examining the position-space compressible energy density at these times, we find that
 the sound pulses which shed behind the obstacle produce a signal higher than any other compressible energy source. These features are approximately $1-3\ \xi$ in size, consistent with the location of the kink point at $k\xi\simeq 2$. 

\subsection{Energy Composition}
\begin{figure}[!t]
\includegraphics[width = \columnwidth]{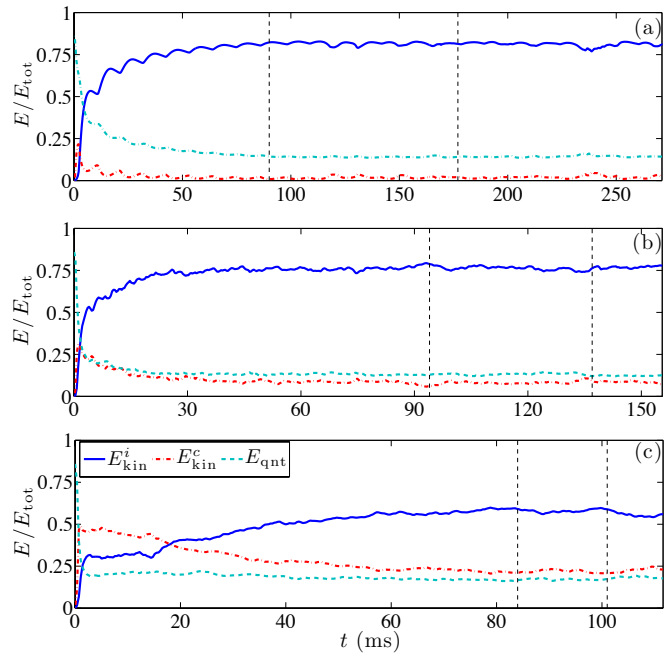} 
\caption{Fractional energy composition as a function of time for the (a) Dipole regime, (b) Cluster regime and (c) Oblique soliton regime. The vertical dashed lines indicate the times at which we have presented the density profiles in Figure \ref{fig2} and the spectra in Figures \ref{fig3} and \ref{fig4}.  One orbital stirring period takes 88 ms for v$=0.4\,c$, 52 ms for v$=0.7\,c$, and 27 ms for v$=1.3\,c$. }
\label{fig5}
\end{figure}

In each stirring regime the system reaches an approximate steady state due to the balance of forcing and damping. However, due to the cyclic nature of the stirring, the total energy continues to exhibit significant fluctuations (of order 10-20\%). In Figure \ref{fig5} we show the fractional kinetic energies $E/E_{\rm{tot}}$ where $E$ is either the total incompressible ($E^i_{\rm kin}$) or compressible ($E^c_{\rm kin}$) kinetic energy, or the total quantum pressure ($E_{\rm qnt}$)~\cite{Numasato10a}. The quantum pressure arises due to sharp variations in the atom density, such as near a vortex core, and signals a departure from hydrodynamics. We find that the fractional energies stabilize relatively quickly to steady-state values. 

We note some global observations. First, the vertical dashed lines, which indicate the times at which we have presented the spectra, all lie within the steady-state regime of the fractional energy. These times are also longer than one full period of the stirring orbits of each case, given in Figure~\ref{fig5}.  Fluctuations observed in spectra are therefore not a consequence of calculating spectra prior to the steady state being reached.  Also, all cases exhibit a short initial stage in which the compressible energy dominates over the incompressible. This is due to the initial compression of the fluid in front of the obstacle when stirring begins.  
%============================================================== 
%============================================================== 
Figure \ref{fig5}(a) clearly demonstrates the predictable and periodic nature of vortex emission within the dipole regime, evident as periodic oscillations in the incompressible energy. One can also see that the energy is largely incompressible, and that the compressible contribution is negligible. Similarly, in the cluster regime the incompressible energy accounts for the overwhelming majority of the total energy, although there is a slightly larger compressible contribution than in the dipole regime. This suggests that the cluster regime is well approximated as incompressible, and may be regarded as a kind of ideal quantum turbulence regime. 

There is a clear difference in energy distribution between the oblique soliton regime [Figure \ref{fig5}(c)] and the other two cases. The compressible energy remains dominant in the initial stages for a significantly longer period in this regime (approximately 20ms). The compressible energy accounts for a much greater proportion of the total energy, so that it is in fact greater than the quantum pressure contribution, in contrast with the other two cases. However, the incompressible energy is still the major contributor in the steady-state regime.
%============================================================== 
\begin{figure}[!t]
\includegraphics[width = \columnwidth]{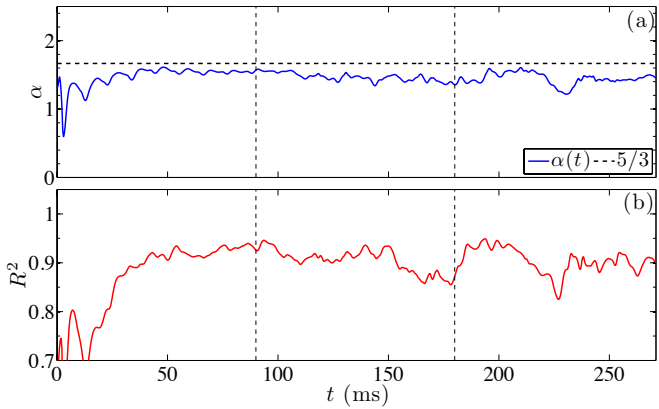} 
\caption{(a) The exponent $\alpha$ ($E^i(k) \propto k^{-\alpha}$) obtained from a least-squares-fit of the infrared spectrum in the dipole regime (A), presented as a function of time. The horizontal dashed line shows the Kolmogorov law $(\alpha = 5/3)$ for comparison. (b) $R^2$ value for the linear fit defined by exponent $\alpha$. The vertical dashed lines indicate each time the stirrer completes a $2\pi$ radian cycle.}
\label{fig5b}
\end{figure}
%============================================================== 

\section{Characterization of Power Laws and Intermittency}\label{sec:fit}

Some of the spectra we have presented in the previous section exhibit intermittent power law behavior. Intermittency is likely to be an irreducible aspect of a trapped system, due to finite size effects. In this section we conduct further, quantitative spectral analysis in order to gain a deeper understanding of this behavior.

For the spectra that exhibit power-law behavior, we perform a linear least-squares-fit in logarithmic space. From this we obtain a best-fit value for the power law exponent, and also an indication of the goodness of fit (given by the $R^2$ value) as functions of time. For a data set $\{y_i\}_{i=1}^N$ with mean $\bar{y}$, and a set of fitted values $f_i$, the $R^2$ value is given by
\begin{equation}\label{R2}
R^2=1-\frac{S_{\rm err}}{S_{\rm tot}},
\end{equation}
where $S_{\rm err}=\sum_{i=1}^N(y_i-f_i)^2$ and $S_{\rm tot}=\sum_{i=1}^N (y_i-\bar{y})^2$.
An $R^2$ value close to 1 indicates the data closely conform to a straight line~\cite{SteelStats}. 
%============================================================== 
\begin{figure}[!t]
\includegraphics[width = \columnwidth]{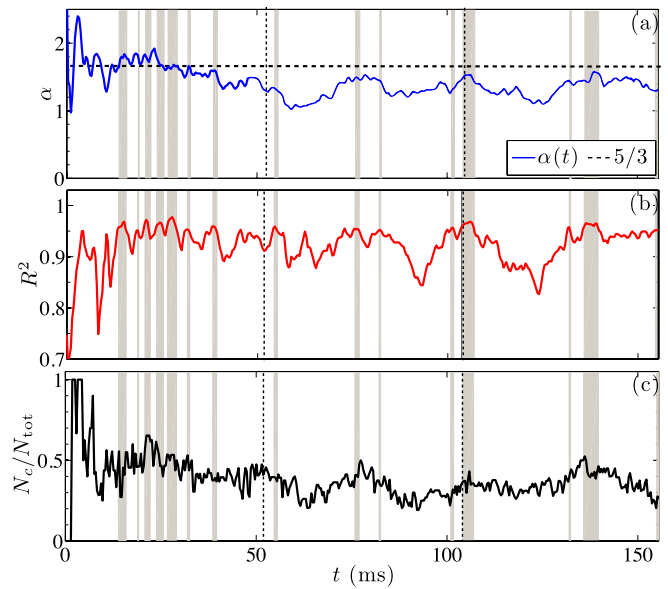} 
\caption{(a) The exponent $\alpha$ ($E^i(k) \propto k^{-\alpha}$) obtained from a least-squares-fit of the infrared spectrum in the cluster regime (B), presented as a function of time. The horizontal dashed line shows the Kolmogorov law $(\alpha = 5/3)$ for comparison. (b) $R^2$ value for the linear fit defined by exponent $\alpha$. (c) Vortex clustered fraction $N_c/N_{\rm tot}$ as a function of time. The vertical dashed lines indicate each time the stirrer completes a $2\pi$ radian cycle. Time intervals where $R^2>0.95$ are shaded grey.}
\label{fig6}
\end{figure}
%============================================================== 
\subsection{Incompressible Spectra}
In the case of the incompressible spectrum, we are only interested in the IR-region (due to universality of the UV region~\cite{Bradley2012a}). We fit within the range $2\pi/30 \leq k\xi \leq 1$, and analyze the dipole and cluster regimes. The incompressible spectrum of the soliton regime does not conform to a power law (as seen in Figure~\ref{fig3})  and we do not consider it further. 

Figure \ref{fig5b} displays the results from this analysis of the dipole regime. Consistent with our observations in Section \ref{sec:enspecA}, we see that the power-law exponent $\alpha$ ($E^i \propto k^{-\alpha}$) for the fit to the data sits near the Kolmogorov value $5/3$ for most of the simulation. However, as previously noted, the spectrum is not particularly linear in log-space compared to the other regimes. The $R^2$ value in the steady state ($t > 90$ms) has a mean of 0.9 and is always less than 0.95. As discussed in Section \ref{sec:regimesA}, the temporal characteristics of the dipole regime are regular and periodic, suggestive of laminar flow. Since we do not expect power-law behavior in the dipole regime, we take $R^2=0.95$ as a benchmark value, such that an $R^2 > 0.95$ demonstrates a goodness of fit exceeding that of the dipole regime.  

The results from the cluster regime are presented in Figure \ref{fig6}. From Figure \ref{fig6}(a) we can see that in the early stages, $\sim 25$ms, the exponent $\alpha$  fluctuates about 5/3 and $R^2 \approx 0.95$. At later times there are points at which $\alpha$ deviates significantly from the Kolmogorov $k^{-5/3}$ law, to values as low as 1. However, the points at which $\alpha$ deviates the most from 5/3 also coincide with decreases in the $R^2$ value of roughly 10\% [Figure 6(b)], indicating that the spectrum is poorly described by a power law at these times. One also observes several short time spans where $\alpha$ returns approximately to 5/3, and $R^2>0.95$, as indicated in Figure \ref{fig6} by the shaded regions.

%============================================================== 
\begin{figure}[!t]
\includegraphics[width = \columnwidth]{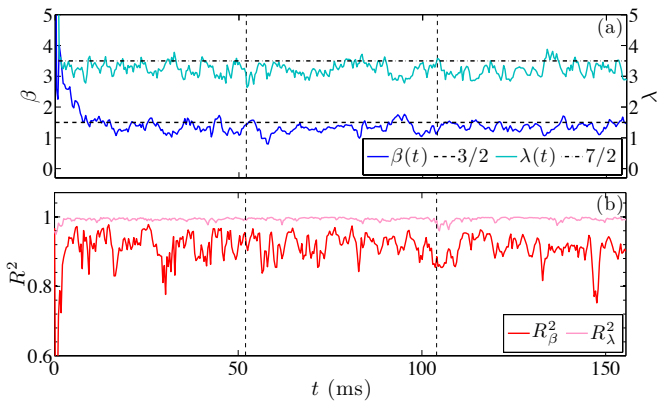} 

\caption{(a) The exponents $\beta$ ($E^c(k) \propto k^{-\beta}, k\xi \lesssim 1$) and $\lambda$ ($E^c(k) \propto k^{-\lambda}, k\xi \gtrsim 1$) obtained from least-squares-fit of the compressible spectrum in the cluster regime (B), presented as functions of time. The horizontal dashed line shows the WWT prediction $(\beta = 3/2)$ for comparison. Also shown for comparison is the dash-dot line $\lambda = 7/2$. (b) $R^2$ value for the best fit analysis as a function of time. The vertical dashed lines indicate each time the stirrer completes a $2\pi$ radian cycle.}
\label{fig7}
\end{figure}
%============================================================== 
To quantify the relation between the spectral linearity in log space and the approach of $\alpha$ to $5/3$, we make use of the correlation between two parameters $X$ and $Y$, defined as
\begin{equation}\label{CXY}
C(X,Y)=\frac{\langle XY\rangle-\langle X\rangle\langle Y\rangle}{\sigma_X\sigma_Y}
\end{equation}
with $\sigma_X^2\equiv \langle X^2\rangle-\langle X\rangle^2$, and for our purposes $\langle X\rangle=N_s^{-1}\sum_{i=1}^{N_s} X(t_i)$ denotes an average over $N_s$ time samples.
We compute the correlation between $|\alpha-5/3|$ and $R^2$ in the steady state (averaging data for times $t_i > 50$ms) and find $C( |\alpha-5/3|,R^2)= -0.63$, indicating a correlation between the approach of $\alpha$ to 5/3 observed in Figure~\ref{fig6} (a) and the approach of the spectrum to a power-law form. 

In Figure \ref{fig6}(c) we have plotted the clustered vortex fraction $N_c/N_{\rm tot}$, where $N_c$ is the number of vortices which have nearest neighbors of the same circulation, and $N_{\rm tot}$ is the total number of vortices.  Vortices detected outside the high density region (farther than $ 0.8R_{\rm{TF}}$ from the trap center) are excluded from the calculation.  Notice that in the early stages ($t \sim 15$ms) where $\alpha$ fluctuates about $5/3$, the clustered fraction is relatively large $N_c/N_{\rm tot}\sim 0.55$. Further comparison between $\alpha$ and the clustered fraction shows that the greatest departures of $\alpha$ from $5/3$ occur when the clustered fraction is reduced to $\sim 0.2$, indicating that vortex dipoles are dominating the flow characteristics. Inspection of the figure shows that when $\alpha$ closely approaches $5/3$, the clustered fraction approaches $\sim 0.5$. In the steady state we find $C( |\alpha-5/3|,N_c/N_{\rm tot})= -0.48$, indicating a notable correlation between the approach of $\alpha$ to $5/3$, and the size of the clustered fraction. Note that the shaded regions in figure \ref{fig6}(c) which indicate that $R^2>0.95$ also approximately coincide with the peaks in the clustered fraction.
\subsection{Compressible Spectra}

%============================================================== 
\begin{figure}[!t]
\includegraphics[width = \columnwidth]{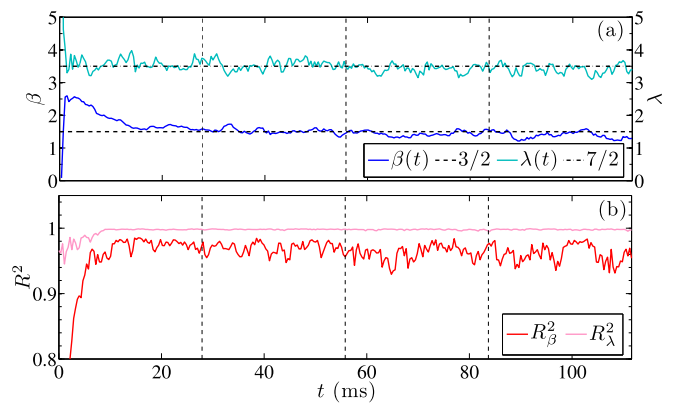} 
\caption{(a) The exponents $\beta$ ($E^c(k) \propto k^{-\beta}, k\xi \lesssim 1$) and $\lambda$ ($E^c(k) \propto k^{-\lambda}, k\xi \gtrsim 1$) obtained from least-squares fit of the compressible spectrum in the soliton regime (C), presented as functions of time. The horizontal dashed line shows the WWT prediction $(\beta = 3/2)$ for comparison. Also shown for comparison is the dash-dot line $\lambda = 7/2$. (b) $R^2$ value for the best fit analysis as a function of time. The vertical dashed lines indicate each time the stirrer completes a $2\pi$ radian cycle.  }
\label{fig8}
\end{figure}
%============================================================== 
We also perform a least-squares fit on the compressible spectrum. We perform this analysis on all three regimes for the UV-region, as all exhibit some indication of power law with exponent $-7/2$. Additionally the IR regions of the cluster and soliton regimes are analyzed, as these regimes exhibit a $-3/2$ power law within this scale range. We fit within the ranges $2\pi/40 \leq k\xi \leq 1.5$ (IR) and $2 \leq k\xi \leq 10$ (UV).
 
  \begin{table*}
\begin{tabular}{c | c | c| | c | c | |c | c  }
\hline
Regime & $\alpha $ ($E^i$, IR-region) & $R^2_\alpha $ & $\beta$ ($E^c$,IR-region)& $R^2_\beta$ & $\lambda$ ($E^c$, UV-region)& $R^2_\lambda$ \\
\hline\hline
Dipole & $\sim 5/3$ & $<0.95 $ (always)  &  -- &  -- &   -- &  -- \\
Cluster & $\sim 5/3$ & $ >0.95$ (intermittently)   & $\sim 3/2$ & $ >0.95$ (intermittently) & $\sim7/2$ & $>0.99 $ (intermittently) \\
Soliton & --&  --  & $\sim 3/2$ &$>0.93$ (always, $\bar{R^2_\beta}=0.96)$  &$\sim7/2$ &  $>0.99 $ (always) \\
\hline
 \end{tabular}
 \caption{Stirring regimes and approximate power-law exponents observed in the energy spectrum. Dashes denote the absence of power-law behavior.}
\label{tab1}
 \end{table*}
 
 We note that, in the dipole regime, whenever a dipole recombination event occurs (e.g., at $t \simeq 118$ ms, see Supplemental Movie 1), the best fit value for $\lambda$ ($E^c(k)\propto k^{-\lambda}$) transiently passes through $7/2$, with $R^2_\lambda > 0.99$ indicating a high degree of linearity. Soon after the event ($t \simeq 125$ms), $\lambda \approx 1.5$ and linearity is greatly degraded ($R^2 \approx 0.67$). 
 
The cluster regime results are displayed in Figure \ref{fig7}. Here $\beta$ ($E^c(k)\propto k^{-\beta}$) quickly approaches $3/2$, within approximately $8$ms, and thereafter fluctuates near $3/2$, with fluctuations of order $10-20\%$ of this value. $R^2_{\beta}$ exhibits significant fluctuations, between values of  $0.97$ and $0.8$, indicating intermittency. In the UV-region $\lambda$ ($E^c(k)\propto k^{-\lambda}$) exhibits similar behavior,  sitting near $7/2$ with fluctuations also of order 10-20\%. $R^2_{\lambda}$ is comparatively high: largely above $0.99$, and always greater than $0.95$.

In the soliton regime,  one again sees that $\beta$  quickly approaches $3/2$, within $t \sim 15$ms [Figure \ref{fig8}(a)]. By this time $R^2_{\beta} >0.95$, indicating that the spectrum is well described by a power law. Thereafter $\beta$ exhibits relatively minor fluctuations from the value $3/2$ $(<10\%)$, and $R^2_{\beta}$ nearly always exceeds $0.95$. Similarly, in the UV region $\lambda$ rapidly conforms to $7/2$, and remains within $10\%$ of this value. For $t > 10$ms $R^2_{\lambda}$ is always above $0.99$.
We summarize our results on power law spectra for the dipole, cluster and soliton regimes in Table~\ref{tab1}.

\section{Discussion and Conclusions}
\label{sec:D&C}
\subsection{Dipole Spectra}
Our numerical investigation has uncovered several surprising results that require further discussion. One such result is the resemblance of the incompressible spectrum to the Kolmogorov $k^{-5/3}$ law in the dipole regime. It appears, however, that this has little to do with turbulent phenomena. The temporal emission and spatial vortex distribution characteristics of the system are highly ordered and regular and there is no significant clustering. These features are clearly shown by Figure~\ref{fig5}(a) and Figure \ref{fig2}(a) respectively (and in Supplemental Movie 1). However, particular configurations of dipoles may produce an approximate power law over a short spectral range, via interference~\cite{Bradley2012a}. A dipole produces an incompressible spectrum that is oscillatory in the IR region, with the $k$-space oscillation frequency inversely dependent on the dipole separation scale. A range of dipole scales could smooth out the spectral oscillations. We note also that the range of dipole scales is much larger in the dipole regime than in the soliton regime. Furthermore, during the time interval where the $k^{-5/3}$ region develops, the spacing of positive and negative vortices is actually \emph{increasing} [see Supplemental Movie 1, $t\sim 50-100$ ms]. This is caused by dissipative motion carrying the dipoles toward the condensate boundary. 

The lack of incompressible energy at small $k$ in the soliton regime is consistent with a system dominated by a single small dipole scale. Furthermore, although the exponent for the IR region of the dipole spectrum approximates $-5/3$, the spectra show a relatively low level of linearity (as measured by the $R^2$ value) in log space, compared with the intermittent results of the clustering regime [Figure \ref{fig6}]. It is clear that the resemblance to a Kolmogorov law indicates the need for caution when interpreting spectra, and the danger of relying on a single measure for identifying turbulent states.

\subsection{Clustering and Intermittency}
We also observe strong intermittency of the Kolmogorov $k^{-5/3}$ power-law within the cluster regime. It is evident from Figure \ref{fig5}(b) that compressible energy, which largely originates from dipole recombination, is not a major contributor within this regime. However, dipole recombination can potentially disrupt an inverse cascade \cite{Kobayashi05a}.
It may also be the case that the clusters that are produced do not have sufficient spatial and temporal extent to support a stable power law. Indeed, the stirring obstacle significantly disrupts freely developing vortex flow, inhibiting clusters of size comparable to the stirring radius from forming. This is consistent with our observation that the $k^{-5/3}$ law does not extend to wavenumbers lower than $k = 2\pi/(40\xi)$, where $40\xi$ is the radial obstacle location. In addition, our stirring procedure usually produces clusters of only two, and at most four vortices, whereas the synthetically generated clusters in \cite{Bradley2012a} that produce a very clear power law over a decade of wavenumbers contain more than 10 vortices. It is also evident from Figure \ref{fig6}(c) that, despite the continuous forcing mechanism, dipoles are still the dominant vortex structures as the clustered fraction is below 0.5 for the majority of the simulation. 

\subsection{Weak-Wave Spectra}\label{sec:wws}
The power-law behavior observed in the compressible spectrum of the clustering and soliton regimes is indicative of weak-wave turbulence in the IR-region. We observe a $k^{-3/2}$ compressible energy spectrum; in the presence of a condensate this spectrum indicates a direct cascade of acoustic energy, driven by three-wave interactions~\cite{Nazarenko06a}.

In the UV-region, the origin of the observed $-7/2$ power law is less clear. We note that the dispersion relation is approximately quadratic at high wavenumber; hence the three-wave kinetic equation that yields the $-3/2$ law in the IR-region is not relevant in the UV-region~\cite{Nazarenko06a}. The cross over to the $-7/2$ power law occurs at $k\xi\simeq 2$, suggesting that four-wave interactions are responsible for the transport of energy at larger wavenumbers~\cite{Zakharov1992}. Further analysis is required to identify the origin of this power law in the presence of a BEC.
\subsection{Forcing Scales and Cascades}
We have no direct evidence of energy cascades, largely due to the difficulty in computing unambiguous fluxes of incompressible and compressible components in a compressible superfluid~\cite{Numasato10a}. Furthermore, we do not observe any evidence for spontaneous vortex clustering either spatially or temporally~\cite{ReevesArXiv}.
The approximate $k^{-5/3}$ incompressible spectrum observed in the clustering regime suggests an inertial range for vortex energy, but the direction of any associated cascade is not clear. It has also been noted that dipole recombination can provide a mechanism for a direct energy cascade~\cite{Numasato10a}. However, as observed in Ref.~\cite{Bradley2012a}, if the forcing scale is near $k\xi=1$ and dipole recombination is suppressed, an inverse cascade of energy to larger scales might occur. This is a consequence of the shape of the incompressible spectrum in the UV-region, which has a universal $k^{-3}$ form due to the structure of the vortex core in 2D, and thus is unavailable for dynamical energy transport. The scale of forcing due to vortex dipole creation behind a stirring obstacle is of order $k\xi=1$, as indicated in Figure \ref{fig5c}(a).  The correlation between the clustered fraction and the approach of the incompressible spectrum to a $k^{-5/3}$ power law is consistent with an intermittent inverse-energy cascade.

The WWT power laws are most clearly observed in the soliton regime. 
The $k^{-3/2}$ spectrum corresponds to a direct cascade~\cite{Nazarenko06a}, suggesting acoustic forcing at small wave numbers. The large peak shown in Figure \ref{fig5c}(b) that drifts towards small $k$ in the compressible spectrum is consistent with this interpretation. 
\subsection{Conclusions and Outlook}
To summarize, we have investigated two-dimensional quantum turbulence in Bose-Einstein condensates using damped Gross-Pitaevskii theory. The range of stirring parameters we have explored exhibits a variety of vortex emission regimes, with different temporal characteristics.  A penetrable obstacle ($V_0\lesssim \mu$) moving at sufficient subsonic speeds ($0.3c\lesssim{\textrm v}\lesssim c$) results in the smooth, periodic emission of vortex dipoles. Maintaining a subsonic stirring velocity but increasing the obstacle strength so that it becomes impenetrable results in increasingly sporadic emission of vortices and the production of like-charged vortex clusters. Increasing the stirring speed into the supersonic regime results in the shedding of dark solitons, which decay into chains of vortex dipoles. 

 Analysis of the incompressible kinetic energy spectra shows that the cluster regime intermittently exhibits a Kolmogorov $k^{-5/3}$ power law within the scale range $2\pi/40 < k\xi < 1$ [Figure \ref{fig3}(b)]. The size of the clustered fraction is negatively correlated with the deviation of the power-law exponent from $-5/3$. This regime simultaneously exhibits two intermittent power laws in the compressible energy spectrum with exponents $-3/2$ and $-7/2$. In contrast, the oblique soliton regime  does not exhibit power-law behavior within the IR region of the incompressible spectrum but instead exhibits strong and temporally robust $-3/2$ and $-7/2$ power laws in the compressible spectrum. The infrared power-law ($-3/2$) is consistent with weak-wave turbulence~\cite{Nazarenko06a}, whereas the origin of the $-7/2$ UV power law presents an intriguing avenue for future work
 
 The intermittency of the Kolmogorov $k^{-5/3}$ law in the cluster regime raises questions as to how one can experimentally produce a state of vortex turbulence that is closer to being fully developed than that which we have produced in this work. Identifying forcing that leads to larger clustered fractions would provide a way to further suppress dipole decay. Additionally, the cyclic nature of the stirring mechanism appears to limit the range over which a power law can be observed, and to disrupt clusters at large scales, introducing intermittency. Identifying experimentally realizable stirring schemes that avoid these issues remains a future challenge. 
\section*{Acknowledgements}
We thank Tom Billam for a critical reading of this manuscript, and Sam Rooney for useful discussions.
We are supported by the Royal Society of New Zealand and the Marsden Fund under grants UOO162 and UOO004 (AB), the University of Otago (MR), and the US National Science Foundation grant PHY-0855467 (BA).
%\bibliographystyle{prsty}

%\bibliography{References}

\begin{thebibliography}{10}

\bibitem{Vinen07a}
W.~F. Vinen and R.~J. Donnelly, Physics Today {\bf 60},  43  (2007).

\bibitem{Bar2001.book.Superfluid}
C.~F. Barenghi, R.~J. Donnelly, and W.~F. Vinen, {\em {Quantized vortex
  dynamics and superfluid turbulence}} (Springer, Berlin, New York, 2001).

\bibitem{Berloff02a}
N.~G. Berloff and B.~V. Svistunov, Phys. Rev. A {\bf 66},  013603  (2002).

\bibitem{Tsubota09a}
M. Tsubota, J. Phys-Cond. Mat. {\bf 21},  164207  (2009).

\bibitem{Kozik09a}
E. Kozik and B. Svistunov, J. Low Temp. Phys. {\bf 156},  215  (2009).

\bibitem{Nowak2011a}
B. Nowak, D. Sexty, and T. Gasenzer, Phys. Rev. B {\bf 84},  020506  (2011).

\bibitem{Bewley08b}
G.~P. Bewley, M.~S. Paoletti, K.~R. Sreenivasan, and D.~P. Lathrop, P Natl.
  Acad. Sci. USA {\bf 105},  13707  (2008).

\bibitem{Kolmogorov1941}
A.~N. Kolmogorov, Dokl. Akad. Nauk. SSSR {\bf 30},  301  (1941).

\bibitem{Maurer1998a}
J. Maurer and P. Tabeling, Europhys. Lett. {\bf 43},  29  (1998).

\bibitem{Araki2002}
T. Araki, M. Tsubota, and S.~K. Nemirovskii, Phys. Rev. Lett. {\bf 89},  145301
   (2002).

\bibitem{Kobayashi05a}
M. Kobayashi and M. Tsubota, Phys. Rev. Lett. {\bf 94},  065302  (2005).

\bibitem{Kobayashi2007a}
M. Kobayashi and M. Tsubota, Phys. Rev. A {\bf 76},  045603  (2007).

\bibitem{Kobayashi2008}
M. Kobayashi and M. Tsubota, J. Low Temp. Phys. {\bf 150},  587  (2008).

\bibitem{Barenghi08a}
C.~F. Barenghi, Physica D {\bf 237},  2195  (2008).

\bibitem{Sreenivasan1999a}
K. Sreenivasan, Rev. Mod Phys. {\bf 71},  S383  (1999).

\bibitem{Kra1967.PF10.1417}
R. Kraichnan, Phys. Fluids {\bf 10},  1417  (1967).

\bibitem{Lei1968.PF11.671}
C. Leith, Phys. Fluids {\bf 11},  671  (1968).

\bibitem{Bat1969.PF12.II233}
G. Batchelor, Phys. Fluids {\bf 12},  II  (1969).

\bibitem{Kra1980.RPP5.547}
R.~H. Kraichnan and D. Montgomery, Rep. Prog. Phys. {\bf 43},  547  (1980).

\bibitem{Kellay2002a}
H. Kellay and W. Goldburg, Rep. Prog. Phys. {\bf 65},  845  (2002).

\bibitem{Mon1974.PF6.1139}
D. Montgomery and G. Joyce, Phys. Fluids {\bf 17},  1139  (1974).

\bibitem{Les2008.Turbulence}
M. Lesieur, {\em {Turbulence in Fluids}}, 4th ed. (Kluwer Academic Publishers,
  Netherlands, 1990).

\bibitem{Boffetta12a}
G. Boffetta and R.~E. Ecke, Annu. Rev. Fluid Mech. {\bf 44},  427  (2012).

\bibitem{Rutgers1998}
M. Rutgers, Phys. Rev. Lett. {\bf 81},  2244  (1998).

\bibitem{Proment09a}
D. Proment, S. Nazarenko, and M. Onorato, Phys. Rev. A {\bf 80},  051603
  (2009).

\bibitem{Raman2001}
C. Raman {\it et~al.}, Phys. Rev. Lett. {\bf 87},  210402  (2001).

\bibitem{Sch2004.PRL93.210403}
V. Schweikhard {\it et~al.}, Phys. Rev. Lett. {\bf 93},  210403  (2004).

\bibitem{Henn09a}
E.~A.~L. Henn {\it et~al.}, Phys. Rev. Lett. {\bf 103},  045301  (2009).

\bibitem{Seman2011a}
J.~A. Seman {\it et~al.}, Laser Phys. Lett. {\bf 8},  691  (2011).

\bibitem{Car2012.JLTP2012.49}
M. Caracanhas {\it et~al.}, J. Low Temp. Phys. {\bf 166},  49  (2012).

\bibitem{Neely2012a}
T.~W. Neely {\it et~al.}, {Characteristics of Two-Dimensional Quantum
  Turbulence in a Compressible Superfluid}, arXiv:1204.1102, 2012.

\bibitem{TWNThesis}
T.~W. Neely, Ph.D. thesis, University of Arizona, Tucson Arizona, USA, 2010.

\bibitem{ECSThesis}
E.~C. Samson, Ph.D. thesis, University of Arizona, Tucson Arizona, USA, 2012.

\bibitem{Chu2001a}
H.-C. Chu and G. Williams, Phys. Rev. Lett. {\bf 86},  2585  (2001).

\bibitem{Chu2001b}
H.-C. Chu and G.~A. Williams,  in {\em Quantized vortex dynamics and superfluid
  turbulence}, edited by C.~F. Barenghi, R.~J. Donnelly, and W.~F. Vinen
  (Springer-Verlag, Berlin Heidelberg, 2001), pp.\ 226--232.

\bibitem{Parker05a}
N.~G. Parker and C.~S. Adams, Phys. Rev. Lett. {\bf 95},  145301  (2005).

\bibitem{Nazarenko06a}
S. Nazarenko and M. Onorato, Physica D {\bf 219},  1  (2006).

\bibitem{Wang2007a}
S. Wang, Y.~A. Sergeev, C.~F. Barenghi, and M.~A. Harrison, J. Low Temp. Phys.
  {\bf 149},  65  (2007).

\bibitem{Numasato2009}
R. Numasato and M. Tsubota, Journal of Physics: Conference Series {\bf 150},
  032074  (2009).

\bibitem{Horng09a}
T.~L. Horng {\it et~al.}, Phys. Rev. A {\bf 80},  023618  (2009).

\bibitem{Numasato10a}
R. Numasato and M. Tsubota, J. Low Temp. Phys. {\bf 158},  415  (2010).

\bibitem{Sasaki2010}
K. Sasaki, N. Suzuki, and H. Saito, Phys. Rev. Lett. {\bf 104},  150404
  (2010).

\bibitem{White10a}
A.~C. White {\it et~al.}, Phys. Rev. Lett. {\bf 104},  075301  (2010).

\bibitem{Numasato10b}
R. Numasato, M. Tsubota, and V.~S. L'vov, Phys. Rev. A {\bf 81},  063630
  (2010).

\bibitem{Schole2012a}
J. Schole, B. Nowak, and T. Gasenzer, {Critical Dynamics of a Two-dimensional
  Superfluid near a Non-Thermal Fixed Point}, arXiv:1204.2487, 2012.

\bibitem{Bradley2012a}
A.~S. Bradley and B.~P. Anderson, Energy spectra of vortex distributions in
  two-dimensional quantum turbulence, arXiv:1204.1103, 2012.

\bibitem{White2012a}
A.~C. White, C.~F. Barenghi, and N.~P. Proukakis, Phys. Rev. A {\bf 86},
  013635  (2012).

\bibitem{Zakharov1992}
V. Zakharov, V. L'vov, and G.~E. Falkovich, {\em {Kolmogorov Spectra of
  Turbulence 1: Wave Turbulence}} (Springer-Verlag, New York, 1992).

\bibitem{Dyachenko1992}
S. Dyachenko, A.~C. Newell, A. Pushkarev, and V.~E. Zakharov, Physica D {\bf
  57},  96  (1992).

\bibitem{Neely10a}
T.~W. Neely {\it et~al.}, Phys. Rev. Lett. {\bf 104},  160401  (2010).

\bibitem{Rooney11a}
S.~J. Rooney, P.~B. Blakie, B.~P. Anderson, and A.~S. Bradley, Phys. Rev. A
  {\bf 84},  023637  (2011).

\bibitem{Gardiner03a}
C.~W. Gardiner and M.~J. Davis, J. Phys. B {\bf 36},  4731  (2003).

\bibitem{Bradley08a}
A.~S. Bradley, C.~W. Gardiner, and M.~J. Davis, Phys. Rev. A {\bf 77},  033616
  (2008).

\bibitem{Blakie08a}
P.~B. Blakie {\it et~al.}, Adv. in Phys. {\bf 57},  363  (2008).

\bibitem{Zhou2004}
Q. Zhou and H. Zhai, Phys. Rev. A {\bf 70},  043619  (2004).

\bibitem{NR}
W.~H. Press, B.~P. Flannery, S.~A. Teukolsky, and W.~T. Vetterling, {\em
  Numerical Recipies}, 1st  ed. (Cambridge University Press, The Pitt Building,
  Trumpington Street, Cambridge CB2 1RP, 1986).

\bibitem{Frisch92a}
T. Frisch, Y. Pomeau, and S. Rica, Phys. Rev. Lett. {\bf 69},  1644  (1992).

\bibitem{Nore97a}
C. Nore, M. Abid, and M. Brachet, Phys. Rev. Lett. {\bf 78},  3896  (1997).

\bibitem{SteelStats}
R.~G.~D. Steel and J.~H. Torrie, {\em {Principles and Procedures of
  Statistics}} (McGraw-Hill, New York, 1960).

\bibitem{ReevesArXiv}
M.~T. Reeves, T.~P. Billam, B.~P. Anderson and A.~S. Bradley, 
Inverse energy cascade in forced 2D quantum turbulence,
 arXiv:1209.5824, 2012.

\end{thebibliography}

\end{document}